\documentclass{emulateapj}
\usepackage{color,ulem}

\defcitealias{Greiner+15}{G15}

\slugcomment{}

\shorttitle{Are ultra-long GRBs caused by blue supergiant Collapsars, magnetars, or white dwarf TDEs? }
\shortauthors{Ioka, Hotokezaka \& Piran}

\begin{document}


\title{Are Ultra-long Gamma-Ray Bursts
Caused by Blue Supergiant Collapsars, Newborn Magnetars, or
White Dwarf Tidal Disruption Events?
}


\author{Kunihito Ioka\altaffilmark{1,2,3}}
\email{kunihito.ioka@yukawa.kyoto-u.ac.jp}
\author{Kenta Hotokezaka\altaffilmark{4}}
\author{Tsvi Piran\altaffilmark{4}}


\altaffiltext{1}{Center for Gravitational Physics, Yukawa Institute for Theoretical Physics, Kyoto University, Kyoto 606-8502, Japan}
\altaffiltext{2}{Theory Center, Institute of Particle and Nuclear Studies, KEK, Tsukuba 305-0801, Japan}
\altaffiltext{3}{Department of Particle and Nuclear Physics, SOKENDAI (The Graduate University for Advanced Studies), Tsukuba 305-0801, Japan}
\altaffiltext{4}{Racah Institute of Physics, The Hebrew University of Jerusalem, Jerusalem 91904, Israel}


\begin{abstract}
Ultra-long gamma-ray bursts ({\it ul}GRBs) are a new population 
of GRBs with extreme durations of $\sim 10^{4}$ s.
Leading candidates for their origin are
blue supergiant collapsars, magnetars,
and white dwarf tidal disruption events (WD-TDEs)
caused by massive black holes (BHs).
Recent observations of supernova-like (SN-like) bumps associated 
with {\it ul}GRBs challenged 
both the WD-TDE and the blue supergiant models
because of the detection of SNe
and the absence of hydrogen lines, respectively.
We propose that WD-TDEs can accommodate
the observed SN-like bumps
if the fallback WD matter releases energy into the unbound WD ejecta.
The observed ejecta energy, luminosity, and velocity are 
explained by the gravitational energy, Eddington luminosity,
and escape velocity of the formed accretion disk, respectively.
We also show that the observed X-rays can ionize the ejecta,
eliminating lines.
The SN-like light curves (SN 2011kl) for the {\it ul}GRB 111209A
are consistent with all three models,
although a magnetar model is unnatural
because the spin-down time required to power the SN-like bump
is a hundred times longer than the GRB.
Our results imply that TDEs are a possible energy source 
for SN-like events in general and for {\it ul}GRBs in particular.
\end{abstract}





\section{Introduction}

A rich diversity of gamma-ray burst (GRB) phenomena has been recognized over
the last decade.
Long GRBs are thought to arise from collapsars, i.e.,
the collapse of massive stars \citep{Paczynski98}
associated with jets penetrating the stellar envelope
\citep{MacFadyen_Woosley99}.
Short GRBs ({\it s}GRBs) are most likely caused by mergers of neutron stars 
\citep{Eichler+89}.
Flares of soft gamma-ray repeaters
most likely result from the sudden release 
of energy via magnetic field reconnection \citep[e.g.,][]{TD93,TD95,Ioka01}.
Low-luminosity GRBs are still mysterious but could be produced by
shock or jet breakouts from stellar envelopes or winds
\citep{Kulkarni+98,Toma+07,Waxman+07,Bromberg+11,Nakar15,IC16}.
Besides GRBs,
an unusual class of objects similar to Swift J1644+57 (GRB 110328A)
\citep{Bloom+11,Burrows+11,Levan+11,Zauderer+11}
is considered to be tidal disruption events (TDEs) 
that launch relativistic jets \citep{Bloom+11,KP11}

Recently a new class of GRBs has been identified---ultra-long GRBs
({\it ul}GRBs)
\citep{Gendre+13,Levan+14}.
These include
GRB 101225A
\citep[the so-called Christmas burst;][]{Campana+11,Thone+11},
GRB 111209A,
GRB 121027A,
GRB 130925A \citep[e.g.,][]{Bellm+14,Evans+14,Piro+14},
and possibly GRB 141121A \citep{Cucchiara+15},
which are listed in Table~\ref{tab:ULGRB}.
Additional {\it ul}GRB candidates were recently reported
by \cite{Lien+16}. 
The duration of these {\it ul}GRBs, $\sim 10^4$ s,
is much longer than that of the conventional long GRBs,
while the isotropic energy is comparable.
Since the event rate is also comparable, $\sim 1$ Gpc$^{-3}$ yr$^{-1}$
\citep{Levan+14},
these {\it ul}GRBs are not negligible 
from the point of view of the total energy budget
compared with the long GRBs,
possibly manifesting themselves in the neutrino sky
\citep{MI13}.
But it is still unclear whether these bursts are simply extreme examples 
of the long GRB class \citep{ZhangBB+13} or not \citep{Boer+15,GM15}.

\begin{deluxetable*}{lrrrlrr}
\tablecaption{{\it ul}GRBs}
\tablecolumns{7}
\tablewidth{0pc}
\tabletypesize{\small}
\tablehead{
\colhead{GRB} & \colhead{$z$} & \colhead{$T_{90}$} & \colhead{$E_{\rm iso}$} & \colhead{Late Decay} & \colhead{Light Radii of Hosts} & \colhead{GRB Position} \\
\colhead{} & \colhead{} & \colhead{(s)} & \colhead{(erg)} & \colhead{} & \colhead{(pc)} & \colhead{(pc)}
}
\startdata
GRB 101225A & 0.85 & $>7000$ & $1.2 \times 10^{52}$ & $\cdots$ & $<600$ ($80\%$ LR) & $<150$ \\
GRB 111209A & 0.67 & $>10000$ & $5.2 \times 10^{52}$ & $t^{-1.36 \pm 0.05}$ & $700$ ($80\%$ LR) & $<250$ \\
GRB 121027A & 1.77 & $>6000$ & $7 \times 10^{52}$ & $t^{-1.44 \pm 0.08}$ & $\cdots$ & $\cdots$ \\
GRB 130925A & 0.35 & $\sim 4500$ & $1.5 \times 10^{53}$ & $t^{-1.32}$ ($t>300$ ks) & $2400$ ($50\%$ LR) & $<600$ \\
GRB 141121A & 1.469 & $\sim 1410$ & $8 \times 10^{52}$ & $t^{-2.14\pm 0.34}$ ($t>5$ days) & $\cdots$ & $\cdots$ 
\enddata
\tablecomments{Data from \citet{Levan+14,Horesh+15,Piro+14,Cucchiara+15,Schady+15}.}
\label{tab:ULGRB}
\end{deluxetable*}

Several models have been proposed for the {\it ul}GRBs.
Three representative models are
blue supergiant collapsar \citep{Kashiyama+13,Nakauchi+13},
newborn magnetar \citep[][hereafter G15]{Greiner+15},
and white dwarf TDE (WD-TDE) models 
\citep{Gendre+13,Levan+14,MacLeod+14}\footnote{See \cite{Gao+16}
and \cite{Perets+16} for other models.}.
The blue supergiant model is a simple extension 
of the collapsar model for long GRBs,
in which the ultra-long duration arises due to
the much more extended envelope of the progenitor star,
which leads to a much longer activity of the central engine
as discussed in the context of Population III GRBs \citep{SI10,Nagakura+12}.
In the magnetar model,
a rotating neutron star launches a magnetized jet 
for the duration of its spin-down time
\citep{Usov92,TD93,Wheeler+00,Thompson+04}.
The WD-TDE model is also a natural extension of a model for Swift J1644+57
to slightly shorter duration. This is consistent with 
the orbital time of the most bound debris \citep{Rees88} 
and with the variability on a scale of a few hundred seconds
during the burst \citep{KP11}
if the disrupted star is a WD. 

The blue supergiant collapsar and the WD-TDE models
have been challenged recently by the observations of 
supernova-like (SN-like) bumps associated with {\it ul}GRBs.
\citet{Levan+14} have noticed SN-like bumps in
infrared/optical afterglows $\sim 10$ days after some {\it ul}GRBs.
The luminosities of these SN-like bumps
are between those of SNe Ic associated with long GRBs
and superluminous supernovae (SLSNe) \citep[e.g.,][]{Gal-Yam12},
and they are similar to rapidly rising gap transients \citep{Arcavi+16}.
\citet{Nakauchi+13} interpreted the SN-like bump 
within the blue supergiant model as 
emission from an expanding cocoon, which is energized by a jet
during its propagation through the progenitor star.
However, \citetalias{Greiner+15} have reported
detailed light curves and spectra of SN 2011kl associated with 
the {\it ul}GRB 111209A.
They discarded both the blue supergiant model and the WD-TDE model
because of the absence of hydrogen lines 
and the detection of the SN, respectively.
By elimination, they are led to develop a magnetar model
by modifying that of \cite{KB10} for SLSNe.

We re-examine here the origin of the SN-like bumps associated with
{\it ul}GRBs, 
concluding that the current observations cannot exclude
either WD-TDE or blue supergiants as the progenitors of {\it ul}GRBs.
We suggest that WD-TDEs can produce an SN-like bump
if the fallback WD matter releases its gravitational energy into
the surrounding tidally unbound WD ejecta.
As far as the lack of hydrogen lines is concerned, we note that
the observed X-ray emission that arises from the central engine
could ionize the ejecta, removing hydrogen lines 
as well as carbon and oxygen lines 
from the spectrum of SN-like bumps.

After submitting our paper, we noticed \citet{Leloudas+16},
discussing that the SLSN-like transient ASASSN-15lh could originate from
a TDE from a Kerr black hole (BH).
This may be a normal-star TDE but an analogue of our case.
Observationally, many optical TDEs are discovered by exploring the gap
between SNe and SLSNe \citep[e.g.,][]{Arcavi+14}.
Thus some SN/SLSN-like events would be caused by TDEs.

The organization of this paper is as follows. In Section~\ref{sec:blc},
we examine the bolometric light curve of the SN-like bump 
observed in the {\it ul}GRB 111209A
in order to discriminate whether the energy injection is explosive or continuous
before going into specific models.
In Section~\ref{sec:model},
we reproduce the multi-band light curves of each model
and discuss the implications of the parameters obtained.
We find that the model parameters are somewhat unnatural 
for a magnetar model.
In Section~\ref{sec:spec},
we show that the observed X-ray emission could ionize the ejecta,
erasing lines in the spectrum.
In Section~\ref{sec:host},
we calculate the probability for
the location of {\it ul}GRBs
to be concentrated in the nuclei of host galaxies.
Section~\ref{sec:sum} is devoted to the summary and discussions.

We adopt a redshift $z=0.677$ 
with a luminosity distance $d=4080$ Mpc
for GRB 111209A \citep{Levan+14},
and a standard $\Lambda$CDM cosmology with
$\Omega_\Lambda=0.73$, $\Omega_{\rm m}=0.27$ and $H_0=71$ km s$^{-1}$ Mpc$^{-1}$.

\section{The bolometric light curve of the supernova-like bump}\label{sec:blc}

We begin, in Sec.~\ref{sec:global}, 
with a discussion of the global characteristics of ejecta 
that can produce an SN-like bump.
We obtain an order of magnitude estimate of physical quantities.
We then examine, in Sec.~\ref{sec:injection},
the light curve of the SN-like bump SN 2011kl
associated with {\it ul}GRB 111209A.
Our goal is to determine whether the energy injection is 
explosive (with a short timescale compared to the bump) 
or continuous (with a timescale comparable to the bump),
before discussing specific models in Sec.~\ref{sec:model}.

\subsection{Global characteristics of the ejecta}\label{sec:global}

The observations at the peak of the SN-like bump
can constrain the physical properties of the ejecta.
The peak time of the bump, $t_p \sim 10$ days, 
is much longer than the {\it ul}GRB 111209A.
It can be identified as the diffusion timescale of photons in the ejecta:
\begin{eqnarray}
t_d &=& \sqrt{\frac{3\kappa M}{4\pi v c}}
\nonumber\\
&\sim & 15\,{\rm day}
\left(\frac{\kappa}{0.1\,{\rm cm}^2\,{\rm g}^{-1}}\right)^{1/2}
\left(\frac{M}{1 M_\odot}\right)^{1/2}
\left(\frac{v}{10^9\,{\rm cm}\,{\rm s}^{-1}}\right)^{-1/2},
\label{eq:td}
\end{eqnarray}
where $M$, $\kappa$, and $v (\ll c,\ {\rm the\ speed\ of\ light})$ are
the mass, opacity, and velocity of the ejecta
\citep{Arnett79,Arnett80,DK13}.
The photospheric velocity is estimated from the peak temperature $T$ 
and luminosity $L$ as
\begin{eqnarray}
v\sim \sqrt{\frac{L}{4\pi t_p^2 \sigma T^4}}
\sim 2 \times 10^9\,{\rm cm}\,{\rm s}^{-1}
\left(\frac{T}{10^4\,{\rm K}}\right)^{-2},
\label{eq:v}
\end{eqnarray}
where $L \sim 3\times 10^{43}\,{\rm erg}\,{\rm s}^{-1}$ is 
the observed bolometric luminosity \citepalias{Greiner+15}
and $\sigma$ 
is the Stefan--Boltzmann constant.
The spectrum observed by X-shooter shows a rest-frame peak at $3000$ \AA\
\citepalias{Greiner+15},
implying a blackbody temperature of $\sim 10^4$ K or higher.\footnote{
Shorter wavelengths than the peak are absorbed by metal lines,
so that the thermal peak may be bluer than the observed one.
\citetalias{Greiner+15} also estimate that the photospheric velocity 
is larger than $2\times 10^{4}$ km s$^{-1}$
by using the width of the absorption line,
although the lines are useless if the elements are ionized
as discussed in Section~\ref{sec:spec}.}
On the other hand, the kinetic energy,
\begin{eqnarray}
E_k=\frac{1}{2} M v^2 \sim 1 \times 10^{51}\, {\rm erg}
\left(\frac{M}{1 M_\odot}\right)
\left(\frac{v}{10^9\,{\rm cm}\,{\rm s}^{-1}}\right)^2,
\label{eq:Ek}
\end{eqnarray}
should be larger than
the total radiated energy 
$\sim L t_p \sim 3\times 10^{49}\, {\rm erg}$
since the radiation pressure accelerates the ejecta
during the optically thick regime.
Thus the mass and velocity of the ejecta are narrowed down to
$0.3 \lesssim M \lesssim 3 M_\odot$ and 
$3 \times 10^{8} \lesssim v \lesssim 2 \times 10^{9}$ cm s$^{-1}$,
respectively.\footnote{
The ejecta mass may be much smaller than the above estimate
if we only require that the ejecta reprocesses injected radiation
into a lower frequency
\citep{Kisaka+15}.
In this case the peak time of the light curve is determined by that of the energy injection, not by the diffusion.
}

\subsection{Energy injection: explosive vs. continuous}\label{sec:injection}

A crucial question is whether the energy injection into the ejecta is
explosive (with a shorter timescale than the peak time)
or continuous (with a timescale comparable to the peak time).
The blue supergiant model is 
an example of explosive injection,
while continuous injection models include the magnetar and WD-TDE models
(see Sec.~\ref{sec:model} for details).

The difference in the energy injection history 
shows up in the light curve of the SN-like bump.
The light curve is obtained by the diffusion equation.
With a one-zone approximation, this equation is 
\begin{eqnarray}
\frac{L(t)}{4\pi R^2} \approx \frac{c}{3\kappa \rho} \frac{E_{\rm int}/V}{R},
\end{eqnarray}
where $R$, $V=4\pi R^3/3$, and $\rho = M/V$
are the radius, volume, and density of the ejecta
\citep{Arnett79,Arnett80,DK13}.
The internal energy of the ejecta $E_{\rm int}$ depends on
adiabatic losses due to the expansion,
an energy injection (with a rate $H(t)$),
and emission (with a luminosity $L(t)$):
\begin{eqnarray}
\frac{d E_{\rm int}}{d t}
=-\frac{E_{\rm int}}{t}+H(t)-L(t),
\label{eq:Eint}
\end{eqnarray}
where we assume that the pressure is radiation-dominated.
The corresponding luminosity is
\begin{eqnarray}
L(t)& = &\frac{t_i}{t_d} \frac{E_{i}}{t_{d}}e^{-t^2/2t_d^2} \nonumber
\\ & & +\frac{e^{-t^2/2t_d^2}}{t_d^2}
\int_0^t dt' t' e^{t'^2/2t_d^2} H(t'),\label{lumi} 
\label{eq:L(t)}
\end{eqnarray}
where $E_{i}$ is an initial internal energy injected at time $t_i$
into an extended ejecta with a radius $vt_{i}$.
An SN-like bump can be produced by either the first 
term, the explosive energy injection,
or the second term, the continuous energy injection.

A wide class of profiles of continuous energy injection can be expressed by
\begin{eqnarray}
H(t)=\frac{E}{t_e}\frac{\ell-1}{(1+t/t_e)^{\ell}},
\label{eq:H(t)}
\end{eqnarray}
where $E$ is the total injected energy,
$t_e$ is the injection timescale,
and $\ell$ is the decay index.
We describe
continuous injection models by $t_e \gtrsim t_d$ or $\ell \le 2$.
We use $\ell=2$ for a magnetar model and $\ell=5/3$ for a TDE model.

\begin{figure*}
  \begin{center}
    \includegraphics[width=170mm]{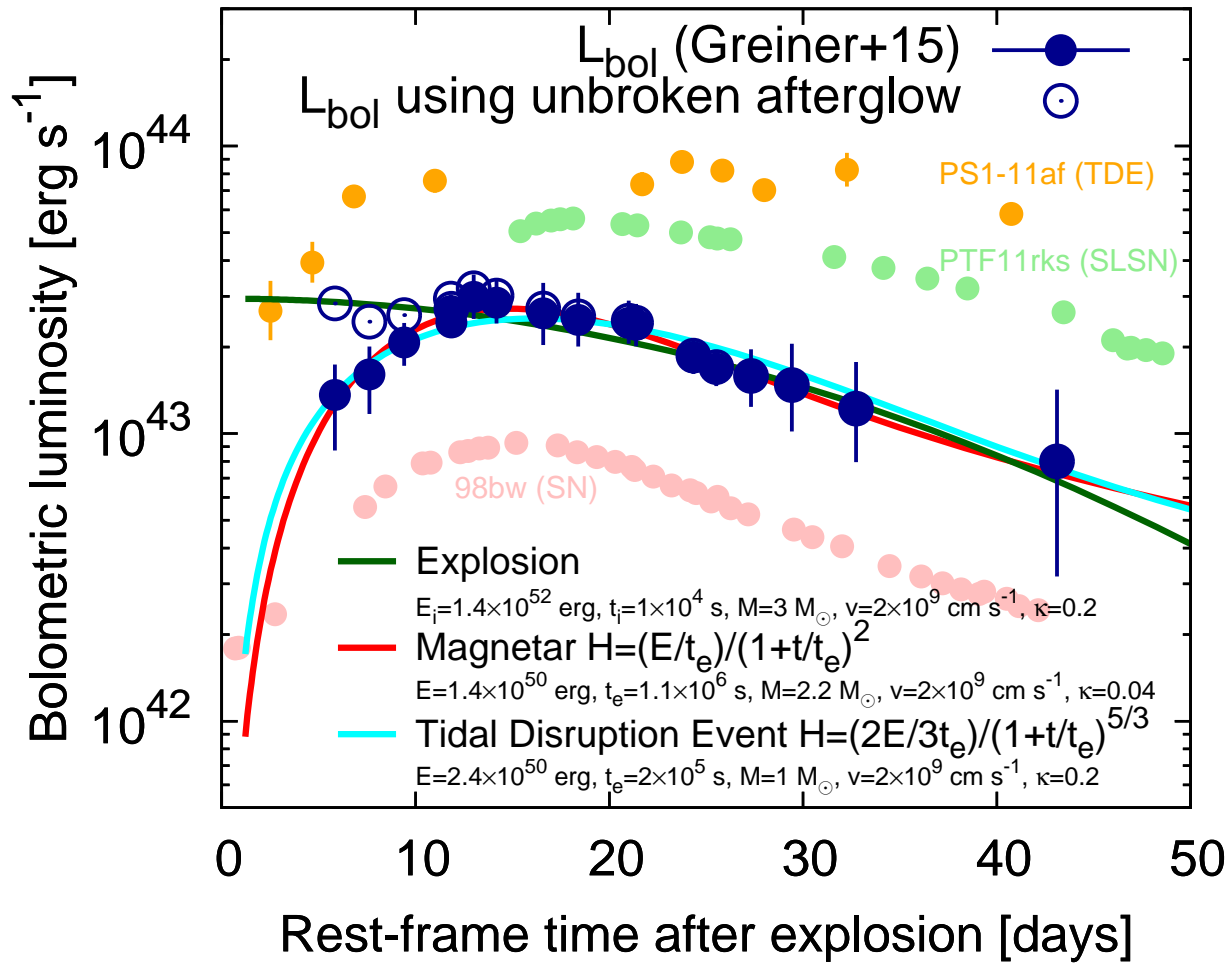}
    \caption{Bolometric light curve of the SN-like bump (SN 2011kl)
      associated with the {\it ul}GRB 111209A 
      with $1\sigma$ error bars ({\it blue}),
      integrated over rest-frame wavelengths of 230--800 mm,
      with the broken afterglow in Equation~(\ref{eq:brokenAG})
      and host galaxy components subtracted from the total light curve
      \citepalias{Greiner+15}.
      We also show the same bolometric light curve by using
      the unbroken afterglow in Equation~(\ref{eq:unbrokenAG}) instead of 
      the broken afterglow ({\it open circle}).
      The former light curve favors a continuous energy injection
      such as the magnetar model ({\it red curve};
      $H=\frac{E/t_e}{[1+(t/t_e)]^2}$ with 
      $E=1.4\times 10^{50}$ erg, $E_i=0$ erg, 
      $t_e=1.1\times 10^6$ s, $M=2.2M_\odot$,
      $v=2\times 10^{9}$ cm s$^{-1}$, $\kappa=0.04$)
      and the TDE model ({\it cyan curve};
      $H=\frac{2E/3t_e}{[1+(t/t_e)]^{5/3}}$ with 
      $E=2.4\times 10^{50}$ erg, $E_i=0$ erg,
      $t_e=2\times 10^5$ s, $M=1M_\odot$,
      $v=2\times 10^{9}$ cm s$^{-1}$, $\kappa=0.2$), while
      the latter one prefers explosive energy injection
      such as the blue supergiant model ({\it green curve};
      the first term in Equation~(\ref{eq:L(t)}) with 
      $E_i=1.4\times 10^{52}$ erg, $E=0$ erg, 
      $t_i=1\times 10^4$ s, $M=3M_\odot$,
      $v=2\times 10^{9}$ cm s$^{-1}$, $\kappa=0.2$).
      For comparison, we also plot
      SN 1998bw/GRB 980425 \citep{Galama+98},
      SLSN PTF11rks \citep{Inserra+13},
      and TDE PS1-11af \citep{Chornock+14}.
}
    \label{fig:Lbol}
  \end{center}
\end{figure*}

In principle, the above parameters can be derived from  
the observed bolometric light curve.
Explosive injection models show a characteristic light curve
with time dependence $\propto e^{-t^2/2t_d^2}$ 
as in the first term of Equation (\ref{eq:L(t)}), i.e.,
the light curve is flat up to the diffusion time $t_d$
and then it decays exponentially.\footnote{The shape of the light curve depends 
weakly on the density profile of the ejecta, which is not taken into
account in our one-zone analysis 
(see, e.g., \citealt{Rabinak+11} for discussions).
In Figure~\ref{fig:explosion}, the multi-band light curves are not flat
because the temperature is changing.}
On the other hand, continuous injection models yield a rising light curve 
up to the diffusion time $t_d$ and a power-law decay afterward.
Figure~\ref{fig:Lbol} depicts several examples 
that will be discussed in detail in Section~\ref{sec:model}.
It is worth noting, however, that
the bolometric light curve around the peak ($t\approx t_d$) is insensitive
to the time dependence of the energy injection
because of the convolution with a Gaussian function 
in Equation~(\ref{lumi}).
Therefore, there is a significant degeneracy among the models 
if the observed bolometric light curve is known only around the peak.

\begin{figure}
  \begin{center}
    \includegraphics[width=85mm]{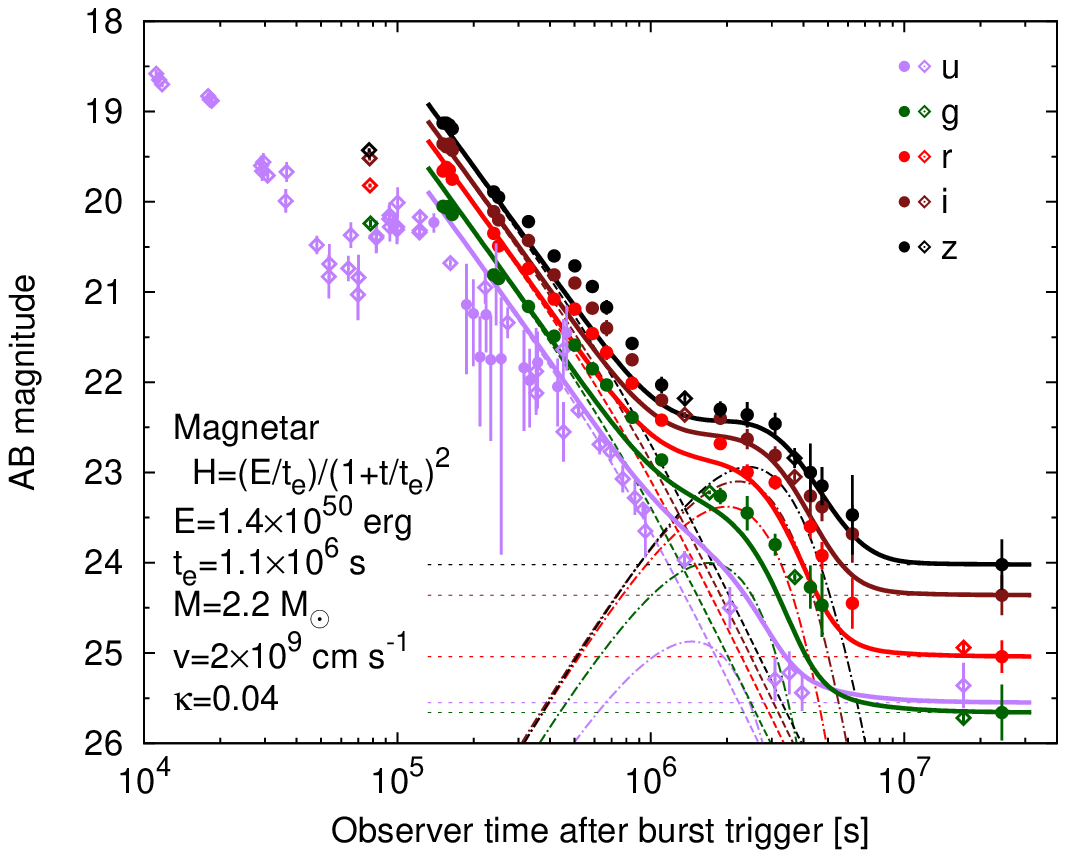}
    \caption{The optical/near-infrared $u$, $g$, $r$, $i$, $z$-band 
      light curves of the {\it ul}GRB 111209A by using a magnetar model 
      in Equations (\ref{eq:L(t)}), (\ref{eq:H(t)}), and (\ref{eq:T})
      with the decay index $\ell=2$, 
      the total injected energy $E=1.4 \times 10^{50}$ erg, 
      $E_i=0$ erg,
      the energy injection time $t_e=1.1 \times 10^6$ s,
      the ejecta mass $M=2.2 M_\odot$,
      the ejecta velocity $v=2 \times 10^9$ cm s$^{-1}$,
      and the opacity $\kappa=0.04$.
      A broken afterglow in Equation (\ref{eq:brokenAG}) is utilized.
      The constant host galaxy contribution 
      is accurately determined at late times.
      Data points are
      $g'$, $r'$, $i'$, $z'$-band observations with GROND 
      in Extended Data Table 1 of \citetalias{Greiner+15},
      $u$-band observations with {\it Swift}/UVOT
      in Extended Data Table 2 of \citetalias{Greiner+15},
      $U$ and White-band observations with {\it Swift}/UVOT
      in Table 4 of \citet{Levan+14},
      and ground-based $u$, $g$, $r$, $i$, $z$-band observations
      in Table 8 of \citet{Levan+14}.
      The magnetar model reproduces the multi-band observations
      within the fluctuations and errors of the flux.}
    \label{fig:magnetar}
  \end{center}
\end{figure}

\begin{figure}
  \begin{center}
    \includegraphics[width=85mm]{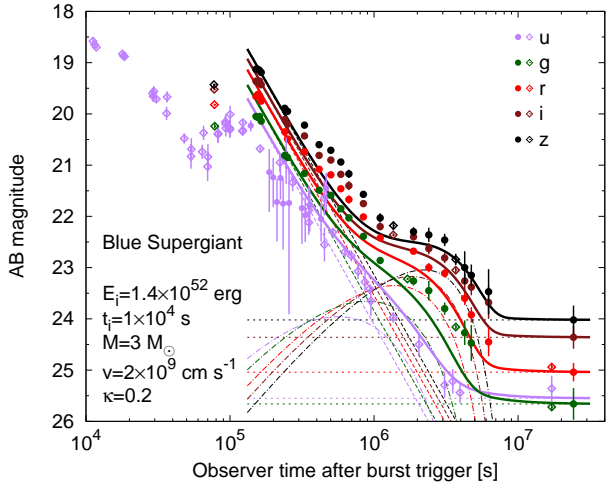}
    \caption{
      The same as Figure~\ref{fig:magnetar} except for
      the use of a blue supergiant model with
      $E_i=1.4 \times 10^{52}$ erg, $E=0$ erg,
      $t_i=1 \times 10^4$ s,
      $M=3 M_\odot$,
      $v=2 \times 10^9$ cm s$^{-1}$,
      and $\kappa=0.2$ in the first term of Equation (\ref{eq:L(t)})
      and Equation (\ref{eq:T}),
      and an unbroken afterglow in Equation (\ref{eq:unbrokenAG}).
      The blue supergiant model also reproduces the multi-band observations
      within the fluctuations and errors of the flux.
      For the data points, see the caption of Figure~\ref{fig:magnetar}.
}
    \label{fig:explosion}
  \end{center}
\end{figure}

Figure~\ref{fig:Lbol} shows 
the bolometric light curve of the SN-like bump (SN 2011kl)
associated with the {\it ul}GRB 111209A \citepalias{Greiner+15}.
Apparently, according to \citetalias{Greiner+15}, 
these observations show a rising light curve,
favoring a continuous energy injection as in a magnetar or TDE
rather than an explosive model.
The late-time data have errors that are too large to distinguish
between a power law (corresponding to a continuous injection)
and an exponential decay (corresponding to an explosive injection).
\citetalias{Greiner+15} extract the SN-like bump component
by subtracting the contributions of the afterglow and the host galaxy 
from the total light curve,
as reproduced in Figure~\ref{fig:magnetar} 
(see Section~\ref{sec:magnetar} for details).
\citetalias{Greiner+15} adopt a broken power-law afterglow,
\begin{eqnarray}
F_{\nu} \propto 
\nu^{-\frac{p-1}{2}}
\left[(t/t_0)^{q \lambda_1}+(t/t_0)^{q \lambda_2}\right]^{-1/q},
\label{eq:brokenAG}
\end{eqnarray}
where $\lambda_1=1.55$, $\lambda_2=2.33$, and $t_0=9.12$ d
in the observer frame.
In Figure~\ref{fig:magnetar},
we use $q=3$ (where larger $q$ values give sharper breaks),
and a spectral index $p=2.6$ \citep{Stratta+13}.

\begin{figure}
  \begin{center}
    \includegraphics[width=85mm]{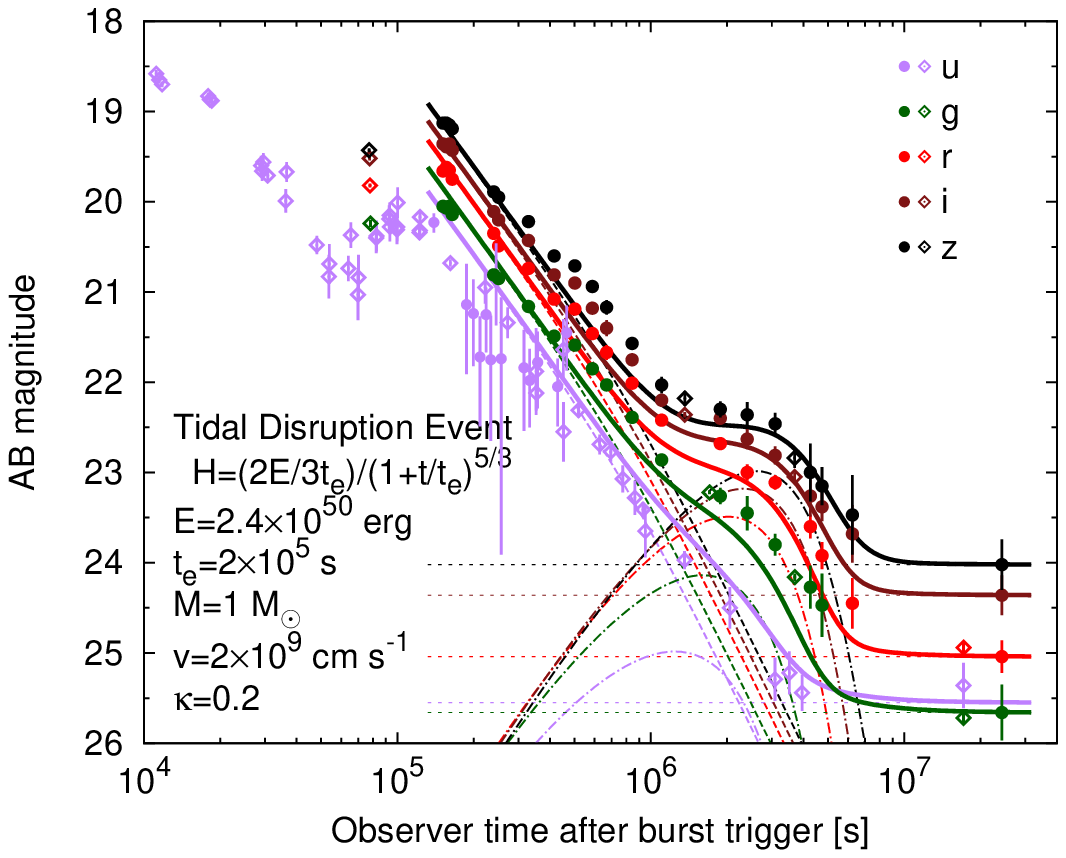}
    \caption{The same as Figure~\ref{fig:magnetar} except for
      the use of a WD-TDE model with
      $\ell=5/3$, 
      $E=2.4 \times 10^{50}$ erg,
      $E_i=0$ erg,
      $t_e=2 \times 10^5$ s,
      $M=1 M_\odot$,
      $v=2 \times 10^9$ cm s$^{-1}$,
      and $\kappa=0.2$ in Equations (\ref{eq:L(t)}), 
      (\ref{eq:H(t)}) and (\ref{eq:T}).
      The same broken afterglow in Equation (\ref{eq:brokenAG}) is utilized 
      as in Figure~\ref{fig:magnetar}.
      The WD-TDE model also reproduces the multi-band observations
      within the fluctuations and errors of the flux.
      For the data points, see the caption of Figure~\ref{fig:magnetar}.
}
    \label{fig:TDE}
  \end{center}
\end{figure}

However, the conclusion supporting continuous energy injection 
depends on the light curve modeling of the afterglow of the {\it ul}GRB 111209A.
If we adopt a single (unbroken) power law for the optical afterglow,
\begin{eqnarray}
F_{\nu} \propto 
\nu^{-\frac{p-1}{2}}
(t/t_0)^{\lambda_3},
\label{eq:unbrokenAG}
\end{eqnarray}
the bolometric light curve will be initially flat,
as shown by open circles\footnote{
To obtain the open circles in Figure~\ref{fig:Lbol}, 
we add the difference between 
Equations (\ref{eq:brokenAG}) and (\ref{eq:unbrokenAG}) to
the bolometric light curve of \citetalias{Greiner+15}.
}
in Figure~\ref{fig:Lbol}.
This instead favors an explosive injection
rather than a continuous one.
Here we use $\lambda_3=2$, and $0.7$ mag fainter normalization at $t=t_0$
than in the broken case,
as shown in Figure~\ref{fig:explosion} 
(see Section~\ref{sec:BSG} for details).
This choice of parameters reduces the contribution of the afterglow
at rest-frame $t \lesssim 10$ days
and therefore 
enhances the contribution of the SN-like signal
during the initial phase.
Although \citetalias{Greiner+15} mention that
``the afterglow light curve shows clear evidence for a steeper
afterglow decay at $>10$ days post-burst, particularly
in the $u'$-band, 
where there is essentially no contribution from the supernova''
(see also \cite{Kann+16}),
such a break is apparently buried in fluctuations and errors of the flux,
and the presence of a break in the $u'$-band
seems unclear in Figures~\ref{fig:magnetar} 
and \ref{fig:explosion}
(see Section~\ref{sec:model}
for implications of the X-ray light curve).
Moreover, the optical afterglow has a large brightening at $\sim 1$ day,
implying the existence of significant fluctuations in flux.

In summary, the bolometric light curve of the SN-like bump
in {\it ul}GRB 111209A does not provide conclusive evidence
as to whether the energy injection is explosive or continuous
because the initial light curve may be
either rising or flat depending on the afterglow modeling,
and the exact shape of the late decay
has large errors.

As shown in Section~\ref{sec:BSG},
the energy required in an explosive injection
is $E_i \sim 10^{52} (v t_i/ 2 \times 10^{13}\,{\rm cm})^{-1}$ erg,
most of which is adiabatically cooled.
This inefficiency limits the size of the progenitor to $v t_i > 10^{13}$ cm,
suggesting a supergiant.

\section{Multi-band light curves of the supernova-like bumps}\label{sec:model}

We consider now 
specific models for {\it ul}GRBs:
a blue supergiant collapsar,
a magnetar,
and a WD-TDE.
We reproduce the expected optical/near-infrared light curves 
of the SN-like bump associated with the {\it ul}GRB 111209A  
in  Figures~\ref{fig:explosion}, \ref{fig:magnetar}, and \ref{fig:TDE},
respectively.

The multi-band light curves are calculated using 
Equations (\ref{eq:L(t)}) and (\ref{eq:H(t)}),
with parameters $(\ell, E, t_e, M, v, \kappa, E_i=0)$ 
or $(E_i, t_i, M, v, \kappa, E=0)$
that are  specified for each model below,
and by multiplying the bolometric flux by the black-body factor
$\pi B_{\nu}/\sigma T^4$ where
\begin{eqnarray}
T=\left(\frac{L}{4\pi \sigma v^2 t^2}\right)^{1/4},\quad
B_{\nu}=\frac{2h \nu^3/c^2}{\exp(h\nu/k_BT)-1}.
\label{eq:T}
\end{eqnarray}
We then convert the flux $f_{\nu}$ [erg s$^{-1}$ cm$^{-2}$ Hz$^{-1}$] to
the AB magnitude $m_{AB}$ using $m_{AB}=-\frac{5}{2} \log_{10} f_{\nu} -48.6$.
We adopt $A_{\rm V}^{\rm Gal}=0.06$ mag Galactic foreground extinction\footnote{
Following \citetalias{Greiner+15},
$A_{u}=0.085$ mag, $A_{g'}=0.066$ mag, $A_{r'}=0.046$ mag, 
$A_{i'}=0.034$ mag, and $A_{z'}=0.025$ mag.
}
and the host galaxy extinction
with rest-frame $A_{\rm V}^{\rm Host}=0.12$ mag
assuming SMC-type dust \citep{Pei92}.\footnote{
The host galaxy extinction at a wavelength $\lambda$ is given by
$A_\lambda=A_V^{\rm Host}+\frac{E_{\lambda-V}}{E_{B-V}} E_{B-V}$
where $\frac{E_{\lambda-V}}{E_{B-V}}$ is given by Equation (5)
in \citet{Pei92}
and $E_{B-V}=A_V^{\rm Host}/R_V$ with $R_V$ in Table 2 of \citet{Pei92}.
}

As discussed in Section~\ref{sec:injection},
the strong afterglow at early times
makes it difficult to distinguish 
between an explosive and a continuous energy injection.
The X-ray afterglow observed by {\it Swift}/XRT
does not show a jet break until $\sim 2 \times 10^{6}$ s 
\citep{Gendre+13,Levan+14},
implying a single power law as in Equation (\ref{eq:unbrokenAG}).
If we use the standard afterglow theory \citep{Sari+98,Nakauchi+13,Stratta+13},
the SN-like component has a flat initial phase (see Figure~\ref{fig:Lbol}),
suggesting an explosive energy injection like a blue supergiant collapser.
To obtain an initially rising light curve,
we must use a broken power law as in Equation (\ref{eq:brokenAG}),
which requires another source for the X-ray emission.
This is possible if the central engine activity is long-lasting,
at least until $\sim 2 \times 10^{6}$ s with a single power law.

\subsection{A blue supergiant Collapsar}\label{sec:BSG}

In a blue supergiant collapsar 
a relativistic jet is launched from an accreting BH formed during the
gravitational collapse of a blue supergiant progenitor
\citep{Kashiyama+13,Nakauchi+13}.
The duration is much longer than that of canonical GRBs
because the radius of a blue supergiant
is much larger than that of an envelope-less star such as a Wolf--Rayet star,
the typical long GRB progenitor \citep{SI10}.
While crossing the stellar envelope,
the jet injects energy into the shocked matter,
producing a hot cocoon surrounding it 
\citep{Begelman+89,Ramirez-Ruiz+02,Matzner03,Bromberg+11b}.
Since the envelope is large, a significant amount of energy, of the order 
of the prompt GRB energy, is dissipated during this phase and is injected into 
the cocoon.
After the jet breaks out, the hot cocoon expands. 
It transfers most of the energy to kinetic energy and the rest is released 
once the optical depth drops. This 
produces an SN-like component similarly to SNe IIP.
The observed SN-like component in GRB 111209A
is brighter than the energetic SN 1998bw/GRB 980425 \citep{Galama+98}
as shown in Figure~\ref{fig:Lbol}.
This may reflect the large envelope of a blue supergiant
and the corresponding large energy of the cocoon.

We adopt the following parameters.
We assume that the  duration of the energy injection into the cocoon, $t_i$,
is comparable to the duration of GRB 111209A.
This is much shorter than the diffusion time $t_d$ in Equation (\ref{eq:td}).
As the energy injection is short-lived,
we use only the first term in Equation~(\ref{eq:L(t)}).
We assume a constant opacity $\kappa=0.2$ cm$^2$ g$^{-1}$.
The true opacity could differ in either direction by a factor of two:
$\kappa \simeq 0.1$ cm$^2$ g$^{-1}$ for singly ionized helium,
$\kappa \simeq 0.2$ cm$^2$ g$^{-1}$ for fully ionized helium,
and $\kappa \simeq 0.4$ cm$^2$ g$^{-1}$ for fully ionized hydrogen.
The remaining three parameters---the total injected energy 
$E_i=1.4 \times 10^{52}$ erg, the ejecta mass $M=3 M_\odot$,
and the ejecta velocity $v=2 \times 10^{9}$ cm s$^{-1}$---are 
chosen to fit the flux, timescale, and temperature (or multi-band data)
of the SN-like bump
(see Equations (\ref{eq:td}) and (\ref{eq:v})).

Figure~\ref{fig:explosion} shows a comparison between the observations
and the model light curves. The blue supergiant collapsar reproduces
the multi-band observations within the fluctuations and errors of the flux,
and it is consistent with the multi-band light curves of GRB 111209A.\footnote{
Our light curve modeling is different from that of \cite{Nakauchi+13},
who  include the large brightening at $\sim 1$ day 
in the model fitting.
We think that this is not appropriate given the GROND data
\citepalias{Greiner+15}.}

A natural feature of this model is that the kinetic energy of the ejecta,
$E_k= 1.2 \times 10^{52}\,{\rm erg}
({M}/{3M_\odot})
({v}/{2 \times 10^{9}\,{\rm cm}\,{\rm s}^{-1}})^2$,
is almost equal to the total injected energy 
$E_i = 1.4 \times 10^{52}$ erg.
In turn, this energy, $E_i$, is  the energy given by the jet to the cocoon
\citep{Bromberg+11b}.
Assuming that the jet luminosity $L_j$ within the envelope
is the same as that of the {\it ul}GRB,
this energy is estimated as \citep{Bromberg+11b}
\begin{eqnarray}
E_i=L_j t_i=L_j T_{90} \left(\frac{t_i}{T_{90}}\right)
=E_{\gamma,{\rm iso}} 
\left(\frac{\theta_j^2}{\epsilon_{\gamma}}\right)
\left(\frac{t_i}{T_{90}}\right),
\label{eq:Ek2}
\end{eqnarray}
where $E_{\gamma,{\rm iso}} \sim 5.2 \times 10^{52}$ erg
is the observed isotropic gamma-ray energy \citep{Levan+14},
the jet duration $t_i$ 
is of the order of $ 2 \times 10^{13}{\rm cm} /0.1 c \approx 6000$ s,
$\theta_j$ ($\gtrsim 0.2$ radian for the absence of a jet break) 
is the opening angle of the jet,
and $\epsilon_{\gamma}$ is the radiative efficiency,
which is of the order of 0.1--0.2. With these parameters 
the factor multiplying $E_{\gamma,{\rm iso}}$ is $\sim 0.2$, 
yielding the required $E_i$.
Note that the two energies, $E_k$ and $E_i$ 
(or $E$ in Equation~(\ref{eq:H(t)})), which are comparable in this case,
are very different in the other models as shown below.

\subsection{A magnetar}\label{sec:magnetar}

When discussing magnetars, 
we have to distinguish between an ``explosive" magnetar
and a ``continuous" magnetar.
An explosive magnetar is one that operates on a short time scale 
(compared to the diffusion time), and in this case 
it is basically the same as the blue supergiant collapsar discussed 
in the previous section. 
To avoid significant adiabatic cooling and an excessive energy budget for the magnetar, it must involve a supergiant progenitor. 
In this case the magnetar could also produce 
the observed ultra-long prompt emission.
The energy injection time can be comparable to the ultra-long prompt timescale
(but this is also longer then and different from the standard central engines 
for GRB magnetars, which are about $100$ s).

A ``continuous" magnetar, i.e., a magnetar of \citetalias{Greiner+15},
is one in which the energy release is comparable to the diffusion time scale. 
In the following we discuss the production of the SN-like bump 
by a continuous magnetar of this kind.

A millisecond magnetar---a highly spinning neutron star with a strong magnetic 
field---has been proposed as a possible central engine of GRBs 
\citep{Usov92,TD93,Wheeler+00,Thompson+04}.
The rotational energy of the neutron star is extracted by 
a relativistic wind of magnetized electron--positron plasma,
which may lead to a relativistic jet for GRBs.
The spin-down time of the dipole emission, $t_s$, is
\begin{eqnarray}
t_{s}&=&\frac{6 I c^3}{B^2 R_{\rm NS}^6 \Omega_i^2}
\nonumber\\
&\sim & 1.6 \times 10^{4}\ {\rm s}
\left(\frac{B}{1 \times 10^{15}\,{\rm G}}\right)^{-2}
\left(\frac{P_i}{2\,{\rm ms}}\right)^2,
\label{eq:ts1}
\end{eqnarray}
where $I$, $R_{\rm NS}$, and $\Omega_i=2\pi/P_i$ are
the moment of inertia, radius,
and angular velocity of a neutron star.
This is adjustable to the required duration of the GRB. 
The total energy is chosen to be larger than the observed released energy.

The model is flexible so as to accommodate, by tuning $P_i$ and $B$, 
diverse transients \citep{Metzger+15,Kashiyama+16}.
In particular, magnetars have been applied to possible central engines
of GRBs on a time scale of tens of seconds or even shorter for {\it s}GRBs
\citep{Usov92,TD93,Wheeler+00,Thompson+04},
including low-luminosity GRBs \citep[e.g.,][]{Mazzali+06,Toma+07},
and the afterglow X-ray plateaus 
on a time scale of a few hours \citep[e.g.,][]{Corsi_Meszaros09}.
Additionally, magnetars have been proposed to power 
SLSNe\footnote{In particular, hydrogen-poor (Type I) SLSNe are difficult 
to explain by radioactivity or interaction with circumstellar material,
and hence the energy injection from the central engine magnetar
to the supernova ejecta has been suggested as the mechanism 
that makes these SLSNe so bright \citep{Gal-Yam12}.}
on a time scale of a few weeks
\citep{KB10,Woosley10,Metzger+15,Mosta+15,Kashiyama+16}.
Although the SN-like bump of the {\it ul}GRB 111209A is fainter than
typical SLSNe by more than one magnitude, and
the spectrum of GRB 111209A is more featureless than that of SLSNe,
a magnetar has been put forward as a possible energy source
for the observed SN-like bump
\citepalias{Greiner+15}.

We adopt the following parameters to calculate the light curve
of the magnetar-powered emission $L(t)$:
the decay index of the energy injection $H(t)$
is determined by the spin-down index $\ell=2$.
We adopt the same ejecta mass $M=2.2 M_\odot$ and 
velocity $v=2\times 10^{9}$ cm s$^{-1}$ as in \citetalias{Greiner+15}.
The remaining three parameters---the energy injection time 
$t_e=1.1 \times 10^{6}$ s,
the total injected energy $E=1.4\times 10^{50}$ erg,
and the opacity $\kappa=0.04$---are chosen to fit the peak time, height, 
and shape (width) of the light curve
(see Equations (\ref{eq:td}) and (\ref{eq:v})).
Note that $t_e$, $E$, and $\kappa$ are not written 
explicitly in \citetalias{Greiner+15},
but these three parameters are fixed by the choice of $M$ and $v$ and
the three observables---the peak time, height, and width of the light curve.

Figure~\ref{fig:magnetar} reproduces the model described by 
\citetalias{Greiner+15} (see also \cite{Kann+16}) and shows
a comparison between the observations and the model light curves.
The model is consistent 
with the multi-band light curves of 
the {\it ul}GRB 111209A within the fluctuations and errors of the flux.

While the light curve can be fitted by this model of energy injection,
the magnetar model as a whole has several problems.
First, there is
a tension between the properties of the prompt emission
and those required for the production of the SN-like bump.
The energy injection time $t_e=1.1 \times 10^{6}$ s
is determined by the peak time of $\sim 10^6$ s of the SN-like bump.
If we identify $t_e$ with the spin-down time of the magnetar, 
it is different from the duration of {\it ul}GRB 111209A 
by two orders of magnitude.
To explain both the prompt emission and the late SN-like bump 
within a magnetar model, we 
require a peculiar behavior: the magnetic field 
should be initially large, leading to a relatively faster decay,
but then it should change
sometime between the prompt emission and the SN-like bump.
This is possible, but it requires an ad hoc assumption 
that does not arise naturally.
Furthermore, it is puzzling why this behavior is observed here and 
not in other cases.
Note that
\cite{Metzger+15} and \cite{Bersten+16}
adopted a spin-down time comparable to the GRB duration of $\sim 10^{4}$ s,
but this choice of parameter makes the bolometric light curve 
decay too rapidly after the peak.
This is exactly the reason why \cite{Bersten+16} require some amount 
of $^{56}$Ni to fit the light curve.
\cite{Cano+16} adopt a spin-down time similar to ours
and introduce a power-law component 
to fit the fast decay phase after the prompt emission,
but in this case they cannot explain the GRB duration by the magnetar activity,
and yet another central engine has to be invoked to explain that.

Second, the kinetic energy $\sim 10^{52}$ erg of the ejecta 
given now by Equation~(\ref{eq:Ek})
is much larger than the total injected energy $E=1.4\times 10^{50}$ erg.
This means that the magnetar releases most of its energy, $\sim 10^{52}$ erg,
at a very early time
and injects a moderate energy $\sim \frac{1}{2}I \Omega_i^2 \sim 10^{50}$ erg
later at $t_e \sim 10^6$ s for the SN-like activity.
This again requires an unnatural behavior of magnetic fields.
Although this behavior is consistent,
a fair fraction of the prompt energy has to be transferred 
to the kinetic energy,  probably requiring an extended envelope 
as in Equation~(\ref{eq:Ek2}).
In addition, the X-ray afterglow does not show a break 
at $\sim 1 \times 10^6$ s, which is
not fully consistent with the required injection time $t_e$.

Third, the required opacity $\kappa \sim 0.04$ is relatively small 
for ordinary ionized plasma, even though the spectra suggest 
a low metal content with $1/4$ of the solar metallicity
\citep[G15;][]{Mazzali+16}.
This problem is also pointed out by \cite{Bersten+16}.
For larger opacity $\kappa \gtrsim 0.1$,
a smaller ejecta mass $M \lesssim 1 M_\odot$ is required 
because $\kappa$ appears as the combination $\kappa M$ 
in the diffusion time $t_d$ in Equation~(\ref{eq:td}),
but such a small ejecta mass is unlikely for a newborn magnetar.

Finally, it is not clear what is the mechanism 
that converts the Poynting flux of the magnetar wind to heat 
that leads to the observed radiation. 
If the Poynting flux just exerts pressure 
on the ejecta, it will just accelerate it.
A fraction of the Poynting energy would be converted into the thermal energy 
through shocks in the ejecta or reconnections of the magnetic fields,
but the process and its efficiency are not known.
The efficiency could be very different from 
that in the cases of radiation or matter
\citep[e.g.,][]{Bromberg+14}.
Note that in blue supergiant collapsars and WD-TDEs 
the initial energy (for collapsars) or the injected energy (for WD-TDEs) 
is transferred directly to the thermal energy.

In summary, the continuous magnetar model could be made consistent 
with the light curves of GRB 111209A.
However, this happens by choosing somewhat unusual model parameters 
and requires either a varying magnetic field, 
with an order-of-magnitude jump, 
between the prompt {\it ul}GRB phase and the SN-like bump phase 
or a different GRB central engine. 
Although it is somewhat strange, clearly one cannot exclude this model.

\subsection{A WD-TDE}\label{sec:TDE}

A TDE of a WD is also a possible origin of {\it ul}GRBs
\citep{Gendre+13,Levan+14,MacLeod+14}.
A star is tidally disrupted  when it passes near a BH \citep{Rees88}.
The disruption occurs at a tidal radius
$R_T \sim (3 M_{\rm BH}/4\pi \rho_*)^{1/3} \sim 4 \times 10^{10}$ cm
for a typical WD density $\rho_* \sim 10^{6}$ g cm$^{-3}$
and a BH mass $M_{\rm BH} \sim 10^5 M_{\odot}$.
The disrupted bound matter is given elliptical trajectories
with large apocenter distances, 
while unbound material moves on hyperbolic orbits.
The most bound matter has a semimajor axis
\begin{eqnarray}
a_{\min} &\sim & \left(\frac{M_{\rm BH}}{M_*}\right)^{1/3} R_T 
\sim 2 \times 10^{12} {\rm cm} \left(\frac{M_{\rm BH}}{10^{5} M_\odot}\right)^{2/3}
\nonumber\\
&& \times
\left(\frac{M_{*}}{1 M_\odot}\right)^{-1/3} \left(\frac{\rho_{*}}{10^{6}\,{\rm g}\,{\rm cm}^{-3}}\right)^{-1/3}
\label{eq:amin}
\end{eqnarray}
for the stellar mass $M_{*} \sim 1 M_\odot$,
with an orbital time
\begin{eqnarray}
t_0 &\sim& 2\pi \sqrt{\frac{a_{\min}^3}{GM_{\rm BH}}} 
\sim 4 \times 10^{3}\ {\rm s}
\left(\frac{M_{\rm BH}}{10^{5} M_\odot}\right)^{1/2}
\nonumber\\
&& \times
\left(\frac{M_{*}}{1 M_\odot}\right)^{-1/2}
\left(\frac{\rho_{*}}{10^{6}\,{\rm g}\,{\rm cm}^{-3}}\right)^{-1/2}.
\label{eq:t0}
\end{eqnarray}
If the BH accretes the fallback matter and launches a jet,
the WD-TDE model can explain the duration of {\it ul}GRBs.
Here a WD is essential because
a regular star provides too long a timescale \citep{KP11},
which would be inconsistent with the observed variability 
of the prompt phase of this event.

WD-TDEs were suggested by \citet{KP11}\footnote{
These events have been suggested also 
as possible progenitors of SNe Ia \citep{LP89,Rosswog+08,Rosswog+09}.}
when Swift J1644+57
was discovered as luminous X-ray flares over several days
\citep{Bloom+11,Burrows+11,Levan+11,Zauderer+11}.
The flux decay $t^{-5/3}$ after $\sim 10^{6}$ s as well as
the location close to the nucleus of a galaxy
indicate a TDE.
The observed variability on a timescale of a few hundred seconds
implied that the disrupted object is a WD and not a regular star.
The latter would imply a minimal variability timescale 
of the order of $10^4$ s.
The super-Eddington luminosity and non-thermal spectrum strongly suggest
that the TDE launches a relativistic jet.
So far several similar events have been discovered such as
Swift J2058+0516 \citep{Cenko+12}
and Swift J1112-8238 \citep{Brown+15}. 
Their different X-ray characteristics as compared with optical TDE candidates 
suggest that their accretion physics is different, 
either due to the different disrupted star or due to a different geometry,
and that in these cases an efficient accretion results in the formation 
of a jet that produces the X-ray emission.

{\it ul}GRBs have several similarities to Swift J1644+57:\footnote{
An optical/IR bump is associated with Swift J1644+57 \citep{Levan+16},
but it is not clear whether this is the same phenomenon as 
the SN-like bump in GRB 111209A or just the afterglow rebrightening.}
\begin{enumerate}
\item The flux decay is close to, if not exactly the same as, $t^{-5/3}$, 
as shown in Table~\ref{tab:ULGRB},
where we note that the decay index is subject to change 
as a result of the choice of the origin of time.
\item The peak luminosities, $\sim 10^{49}$ erg s$^{-1}$, of {\it ul}GRBs 
are higher than those of Swift J1644+57 and Swift J2058+0516, 
but the differences are only one or two orders of magnitude,
much smaller than the diversity of GRB luminosities
\citep[see Figure 15 in][]{Evans+14}.
\item The differences in duration are also at the same level 
as the luminosities
\citep[see Figure 15 in][]{Evans+14}.
\item The locations of {\it ul}GRBs for which data are available 
(GRB 101225A, GRB 111209A, and GRB 130925A) are
consistent with the nuclei of the host galaxies
(see Section~\ref{sec:host} for further details).
\end{enumerate}
In addition, 
theoretical estimates of the WD-TDE rate 
\citep{KP11,Shcherbakov+13,MacLeod+14}
are consistent with the {\it ul}GRB rate
$\sim 1$ Gpc$^{-3}$ yr$^{-1}$ \citep{Levan+14},
taking the numerous uncertainties into account.

\begin{figure}
  \begin{center}
    \includegraphics[width=85mm]{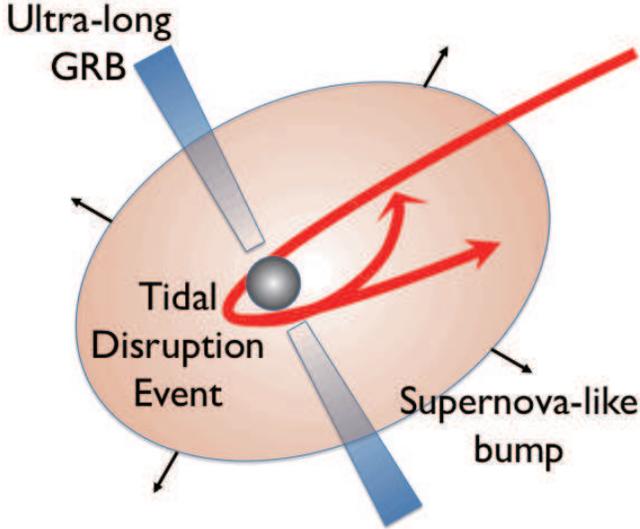}
    \caption{Schematic picture of a TDE
associated with a {\it ul}GRB jet as well as SN-like ejecta.
The SN-like ejecta is powered by 
the gravitational energy of the fallback matter,
which is transferred to the surrounding ejecta via shocks or radiation.
The ejecta releases the injected energy after expansion,
making the SN-like bump like SNe IIP
(see text for details).
}
    \label{fig:TDESN}
  \end{center}
\end{figure}

In view of the observed SN-like bump in GRB 111209A,
\citetalias{Greiner+15} discarded the WD-TDE model.
However, it is premature to do so because
the SN-like bump can be easily powered by 
the material falling back onto the massive BH.
Half of the stellar mass falls back and 
the other half is ejected to infinity (see Figure~\ref{fig:TDESN}).
The fallback mass $\sim M_*/2$ first dissipates its gravitational energy
via shocks between tidal streams
at the outer radius of the most bound orbit,
$\sim a_{\min}$ in Equation~(\ref{eq:amin}),
much further than the tidal radius $R_T$,
as suggested by the optical TDEs \citep{Piran+15}
and simulations \citep{Shiokawa+15}.
The energy dissipated at the radius $a_{\min}$,
\begin{eqnarray}
E &\sim &
\frac{GM_{\rm BH} (M_*/2)}{2 a_{\min}}
\sim 4 \times 10^{51} {\rm erg}
\left( \frac{M_{\rm BH}} {10^{5} M_\odot}\right)^{1/3}
\nonumber\\
&\times & 
\left(\frac{M_*}{1 M_\odot}\right)^{4/3}
\left(\frac{\rho_*}{10^6\,{\rm g}\,{\rm cm}^{-3}}\right)^{1/3},
\label{eq:Edisk}
\end{eqnarray}
is comparable to the energy required for the SN-like bump 
(see Sec.~\ref{sec:global} and below).
Furthermore, an even larger amount of energy can be released
as the matter accretes closer to the BH.
The dissipation lasts for the circularization timescale
of the fallback matter,
which is longer than the orbital time $t_0 \sim 4 \times 10^{3}$ s
given by Equation~(\ref{eq:t0})
by at least a factor of $5$--$10$ \citep{Piran+15,Shiokawa+15}.
While the accretion can be inefficient at this stage \citep{Svirski+15},
the overall accretion rate is super-Eddington. 
If the accretion is efficient, it will lead to 
a strong disk emission in the UV and soft X-rays 
\citep[see][]{Piran+15}\footnote{\cite{Piro+14} and \cite{Bellm+14} 
detected a thermal X-ray component in the {\it ul}GRB 130925A,
but their estimates of the emission size are different,
and it is not clear whether the X-rays are relevant to the disk emission or not.}
or to a powerful outflow via radiation pressure
\citep{Ohsuga+05,SQ09,Metzger_Stone16}.
The emission or outflow is surrounded by the optically thick tidal ejecta,
so that the resulting energy is injected into the ejecta.
The injected energy is radiated later via diffusion and thermalization,
and appears to be an SN-like bump like SNe IIP
(see Figure~\ref{fig:TDESN}).
The luminosity is regulated by the optically thick ejecta
down to the Eddington luminosity of the BH \citep{Shen+16}:
\begin{eqnarray}
L_{\rm Edd} \sim 10^{43}\ {\rm erg}\ {\rm s}^{-1}
\left(\frac{M_{\rm BH}}{10^5 M_\odot}\right),
\label{eq:LEdd}
\end{eqnarray}
which is close to the observed one.
The required ejecta mass $\sim 1 M_\odot$ (see Section~\ref{sec:global})
is consistent within the model uncertainties.
In addition, the escape velocity at $\sim a_{\min}$,
\begin{eqnarray}
v_{\min} &\sim& \sqrt{\frac{GM_{\rm BH}}{2 a_{\min}}}
\sim 2 \times 10^{9}\ {\rm cm}\ {\rm s}^{-1}
\left(\frac{M_{\rm BH}}{10^{5} M_\odot}\right)^{1/6}
\nonumber\\
&& \times
\left(\frac{M_{*}}{1 M_\odot}\right)^{1/6}
\left(\frac{\rho_{*}}{10^{6}\,{\rm g}\,{\rm cm}^{-3}}\right)^{1/6},
\label{eq:vdisk}
\end{eqnarray}
which is of the same order as the velocity of the unbound material,
gives the right value for
the observed ejecta velocity 
(see Sec.~\ref{sec:global} and below).
Therefore a WD-TDE can naturally explain
the SN-like bump observed in GRB 111209A.
Note that the mechanism of the energy injection into the ejecta proposed here
is different from the previous proposals, such as
optical flares from the TDE \citep{Bogdanovic+04,SQ09,CE11}\footnote{
They considered the reprocessed emission by unbound matter,
but only lines, not continuum.
}
and a thermonuclear SN I by tidal compression of the WD
\citep{LP89,Rosswog+08,Rosswog+09,MacLeod+14,MacLeod+16}.
\citet{Metzger_Stone16} 
consider a similar mechanism to ours for optical TDEs,
although the energy is released at the pericenter
by a small fraction of the matter accretion
(not at the apocenter by half of the matter accretion).

We choose the following parameters to calculate the light curve
of the WD-TDE emission $L(t)$:
the decay index of the energy injection $H(t)$
is determined by the fallback index $\ell=5/3$
\citep{Rees88,Phinney89}.
We adopt $\kappa=0.2$ cm$^2$ g$^{-1}$ for fully ionized carbon and oxygen
\citep[e.g.,][]{PM14}.
The mass $M=1 M_\odot$ and 
velocity $v=2\times 10^{9}$ cm s$^{-1}$ 
are taken according to Section~\ref{sec:global}.
The remaining two parameters---the total injected energy 
$E=2.4\times 10^{50}$ erg
and energy injection time $t_e=2 \times 10^{5}$ s---are chosen to 
fit the peak height
and the shape (width) of the light curve.

Figure~\ref{fig:TDE} shows a comparison 
between the observations and the TDE model.
The model reproduces the multi-band light curves of {\it ul}GRB 111209A
within the fluctuations and errors of the flux.

The implications of the required model parameters
for the WD-TDE model are as follows.
First, the energy injection time $t_e=2 \times 10^{5}$ s
is longer than the orbital time $t_0 \sim 4 \times 10^{3}$ s.
This may be natural
because the circularization process of the fallback matter
is rather slow, lasting at least $\sim 5$--$10 t_0$
as found in numerical simulations \citep{Shiokawa+15}
and suggested by optical TDEs \citep{Piran+15}.
{Note that the observed X-ray afterglow most likely arises from a jet,
whose origin is different from the process that produces  that powers the SN-like bump.}
Second, the kinetic energy 
$E_k \sim \frac{1}{2} M v^2 \sim 4 \times 10^{51}$ erg of the ejecta 
is larger than the total injected energy $E=2.4 \times 10^{50}$ erg.
This is natural because the kinetic energy in this case is 
just the original kinetic energy of the unbound stellar debris and 
it is not related to the process that heats the outflow 
at a later time, causing the observed emission.

\section{The spectrum of the supernova-like bump}\label{sec:spec}

The absence of hydrogen lines 
from the SN-like emission of the {\it ul}GRB 111209A is used by 
\citetalias{Greiner+15} to rule out a blue supergiant progenitor.
Carbon and oxygen lines are also not observed,
contrary to the expectation of a WD-TDE \citep{CE11},
although radiation transfer modeling suggests that
the composition is consistent with the carbon-oxygen cores of massive stars
\citep{Mazzali+16}.
In any case, as we will show below, these elements could be ionized 
by radiation, and thus would not display these lines.

The observed X-ray emission at the peak of the SN-like bump
$\sim 10$ days after GRB 111209A
has a luminosity $L_X \sim 10^{44}$ erg s$^{-1}$.
The emission likely comes from the central engine
for the WD-TDE model (see Section~\ref{sec:TDE}).
Although the observed X-ray emission may be an afterglow 
for the blue supergiant model,
the central engine could expose itself at the late time of $\sim 10$ days
and the emission could be comparable to that observed.
The X-rays ionize the ejecta, which is composed,
for example, of oxygen ($Z=8 Z_8$),
at a rate
$t_{\rm ion}^{-1} = n_{\gamma} \sigma_i c$
where $n_{\gamma}=L_X/4\pi R^2 c h \nu$ is the number density of ionizing photons,
$\sigma_i=1 \times 10^{-19} Z_{8}^{-2}$ cm$^{2}$ is the ionization cross section,
and $h\nu=871 Z_8^2$ eV is the ionization energy
\citep{OF06,Metzger+14}.
On the other hand,
the recombination rate is
$t_{\rm rec}^{-1}=n_e \alpha_{\rm rec}$
where $n_e \simeq \rho/2 m_p$ is the electron density,
$\rho=3M/4\pi R^3$ is the density of the ejecta,
and $\alpha_{\rm rec} \sim 2 \times 10^{-11} Z_{8}^2 T_4^{-0.8}$ cm$^{3}$ s$^{-1}$
is the case B recombination coefficient
\citep{OF06,Metzger+14}.
We find that the ionization dominates the recombination,
\begin{eqnarray}
\frac{t_{\rm rec}}{t_{\rm ion}}
&=& \frac{n_\gamma}{n_e}\frac{\sigma_i c}{\alpha_{\rm rec}}
\sim 2 \times 10^2 \,
Z_8^{-6} T_4^{0.8} \left(\frac{L_X}{10^{44}\,{\rm erg}\,{\rm s}^{-1}}\right)
\nonumber\\
&\times & 
\left(\frac{v}{10^{9}\,{\rm cm}\,{\rm s}^{-1}}\right)
\left(\frac{t}{10\,{\rm d}}\right)
\left(\frac{M_{\rm ej}}{1M_\odot}\right)^{-1},
\end{eqnarray}
where the ionization parameter is given by
\begin{eqnarray}
\frac{n_\gamma}{n}
&\sim & \frac{n_\gamma}{n_e/Z}
\sim  9 Z_8^{-1} \left(\frac{L_X}{10^{44}\,{\rm erg}\,{\rm s}^{-1}}\right) 
\left(\frac{v}{10^{9}\,{\rm cm}\,{\rm s}^{-1}}\right)
\nonumber\\
&\times&
\left(\frac{t}{10\,{\rm d}}\right)
\left(\frac{M_{\rm ej}}{1M_\odot}\right)^{-1}.
\end{eqnarray}
Therefore hydrogen, carbon, and oxygen in the ejecta 
could be ionized by X-rays.
The absence of these lines
rejects neither the blue supergiant model nor the WD-TDE model.

As similar examples, hydrogen lines have recently been observed 
in the late-time spectra of a hydrogen-poor SLSN \citep{Yan+15}.
Helium lines have also appeared at a later stage in SN 2008D
\citep{Mazzali+08}.
It is dangerous to conclude the absence of elements from
incomplete observations of lines.

\section{Location in host galaxies}\label{sec:host}

The location in the host galaxies is an important clue to the identification
of the origin of sources.
Association with the nuclei of the host galaxies implies TDEs by massive BHs.
Currently we have astrometric data for GRB 101225A, GRB 111209A, and
GRB 130925A as shown in Table~\ref{tab:ULGRB}.
GRB 101225A and GRB 111209A lie within 150 and 250 pc of the nucleus,
respectively. 
This is consistent with TDEs. 
However, these hosts are compact, 
making it difficult to draw a firm conclusion \citep{Levan+14}.
GRB 130925A is slightly offset from the nucleus of the galaxy 
by $\sim 600$ pc,
while the host galaxy is large with an effective radius of $\sim 2.4$ kpc
\citep{Schady+15}.

The probability of all three bursts accidentally occurring close to the nuclei
may be small, even though each case is not rare in itself.
The probability of occurrence at a position $r<R_i$ in a galaxy with 
a (two-dimensional Gaussian) radius $R_g$
is given by
\begin{eqnarray}
P_i(R_i) = \int_0^{R_i/R_g} r e^{-r^2/2} dr = 1-e^{-R_i^2/2 R_g^2},
\end{eqnarray}
where the radius $R_g$ is related to
$x\%$ light radius $R_{x}$ by $P_i(R_{x})=x\%$.
With Table~\ref{tab:ULGRB}, we obtain the probability
\begin{eqnarray}
P_{101225A} \cdot P_{111209A} \cdot P_{130925A}
&\simeq & 0.0957 \cdot 0.186 \cdot 0.0424
\nonumber\\
&\simeq& 0.000753.
\end{eqnarray}
This is $\simeq 3.4 \sigma$ that 
and could indicate the WD-TDE model.
Note that the compact nature of host galaxies for GRB 101225A and GRB 111209A
also suggests a small mass of nuclear BHs \citep{Levan+14},
which is required for WD-TDEs.

\section{Summary and discussions}\label{sec:sum}

We conclude that WD-TDE can naturally produce {\it ul}GRBs that are
accompanied by SN-like bumps.
The fallback mass has enough gravitational energy to power both the {\it ul}GRB and the SN-like bump.
The WD mass and the velocity of unbound material are consistent 
with the ejecta mass and velocity
required for the observed light curves of the SN-like bumps.
The energy (Equation~(\ref{eq:Edisk}))
and velocity (Equation~(\ref{eq:vdisk}))
for the SN-like bump in the {\it ul}GRB 111209A
are consistent with a picture
in which the fallback matter dissipates energy
at the outer radius of the most bound orbit $\sim a_{\min}$
via mutual shocks between tidal streams. 
Part of the energy is absorbed by the optically thick ejecta,
leading to the SN-like emission at the Eddington luminosity of the BH
in Equation~(\ref{eq:LEdd}) as observed. 

The locations of the bursts  in the centers of their host galaxies are also favorable for a WD-TDE origin.
By combining three events, for which data are available, 
the significance of concentration in nuclei is $\simeq 3.4 \sigma$. 
These locations, of course, do not support 
the magnetar and blue supergiant collapsar models.
The absence of carbon and oxygen lines from the spectrum of the SN-like bump
is not a problem since these elements can be ionized by the observed X-rays.
Together with the flux decay close to $t^{-5/3}$
and a certain similarity to Swift 1644+57,
the WD-TDE model is, to our minds, 
a strong candidate for the origin of {\it ul}GRBs.

The observed SN-like bumps are also still consistent with
the blue supergiant collapsar, 
and more broadly with an explosive injection model.
The light curves of the SN-like bumps are subject to change 
due to the uncertainty of subtraction of the afterglow.
Precise observations of multi-band light curves
are necessary to distinguish 
whether the energy injection is explosive or continuous.
The lack of hydrogen lines is not crucial evidence
against the blue supergiant model
since the observed level of X-rays can ionize hydrogen.

The physical parameters for reproducing the multi-band light curves 
for the magnetar model
(i.e., a magnetar as in \citetalias{Greiner+15}, 
not an explosive, short-lived magnetar; see Section~\ref{sec:magnetar}),
are not attractive.
In particular,
the required spin-down time of the magnetar is much longer than
the prompt emission of {\it ul}GRB 111209A.
The location of {\it ul}GRBs in host galaxies
is also not consistent with that of SLSNe, 
which are possibly powered by magnetars.
Nevertheless the magnetar model is still viable.
Observations of the late-time decay are desirable 
to support or rule out this model. 
A further problem of this model is the need to convert 
the Poynting flux of the magnetar efficiently to heat that can be radiated away.

To conclude, we note that the WD-TDE model provides an interesting connection
between {\it ul}GRBs
and BHs with masses less than $10^5 M_\odot$.
Future observations of {\it ul}GRBs and associated SN-like emission
will probe intermediate-mass BHs if they are WD-TDEs.
Off-axis {\it ul}GRBs might be observed as SNe 
without GRBs in the SN surveys.
Rapidly rising gap transients \citep{Arcavi+16}
are similar to the SN-like bump of {\it ul}GRB 111209A,
and their properties are interesting to study such as by
searches for radio afterglow.

\acknowledgments
We would like to thank J. Greiner 
for providing information about the data used in \citetalias{Greiner+15},
S.~Kisaka for helpful discussions,
and B.~Metzger and an anonymous referee for helpful comments.
This work is supported by 
SOKENDAI (The Graduate University for Advanced Studies),
KAKENHI 24103006, 24000004, 26247042, 26287051 (K.I.), 
by an Israel Space Agency (SELA) grant and the I-Core center for excellence ``Origins" of the CHE-ISF and
by an adv ERC grant TREX. 

\bibliography{ref}

\end{document}